\documentstyle[preprint,eqsecnum,aps]{revtex}
\begin{document}
\draft

\font\Lie=cmff10 scaled\magstep1

\def\d{{d_0}}
\def\O{{\cal O}}
\def\J{{\cal J}}
\def\g{{\cal L}}
\def\Ad{{\hbox{Ad}}}
\def\ad{{\hbox{ad}}}
\def\t#1{{\tilde #1}}
\def\ltimes{\mathbin{\hbox{\hskip1.7pt\vrule height 5.7pt depth -0pt  
                    width .2pt \hskip-1.7pt$\times$}}}                
\def\rtimes{\mathbin{\hbox{$\times$\hskip-1.8pt\vrule height 4.7pt    
                    depth .05pt width .2pt \hskip1.8pt}}}             

\def\Box#1{\mathop{\mkern0.5\thinmuskip
\vbox{\hrule
\hbox{\vrule
\hskip#1
\vrule height#1 width 0pt\vrule}%
\hrule}%
\mkern0.5\thinmuskip}}

\def\beginproof{\noindent{\it Proof.\hskip 0.35em \relax}}
\def\endproof{\hfill$\Box{5pt}$}

\tighten

\preprint{\vbox{
\rightline{ZU-TH-99/2}
\rightline{SU-GP-99/1-1}
\rightline{NSF-ITP-99-08}
\rightline{gr-qc/9902045}}}
\title{A Uniqueness Theorem for Constraint Quantization}
\author{Domenico Giulini}
\address{Theoretical Physics, Z\"urich University, CH-8057 Z\"urich,
         Switzerland}
\address{
Max-Planck-Institut f\"ur Gravitations\-physik,
Schlaatzweg~1,
D--14473 Potsdam,
Germany}
\author{Donald Marolf}
\address{Institute for Theoretical Physics, University of California, 
Santa Barbara, 93106}
\address{
Max-Planck-Institut f\"ur Gravitations\-physik,
Schlaatzweg~1,
D--14473 Potsdam,
Germany}
\address{Physics Department, Syracuse University, Syracuse,
         New York 13244}
\date{October, 1998}
\maketitle
\begin{abstract}
This work addresses certain ambiguities in the Dirac approach to
constrained systems.  Specifically, we investigate the space of
so-called ``rigging maps'' associated with Refined Algebraic
Quantization, a particular realization of the Dirac scheme.
Our main result is to provide a condition under which the rigging
map is unique, in which case we also show that it is given by
group averaging techniques. Our results comprise all cases
where the gauge group is a finite-dimensional Lie group.
\vskip1.0truecm
\noindent
PACS numbers 04.60-m, 04.40.Ds, 02.20.-a
\end{abstract}
\vfil
\eject
\baselineskip = 16pt
\section{Introduction}

Our goal here is to develop and cement the mathematical
structure of the Dirac 
Quantization procedure \cite{Dirac} for constrained systems.
This `procedure' involves introducing the constraints as operators
on some space
and then taking only those states which are annihilated by the 
constraints to be `physical'.  These physical states should then
be made into a (physical) Hilbert space.
Before discussing the details, we
remind the reader that this procedure is expected to be an essential
part of any attempt to quantize gravity using
canonical methods \cite{AAbook,KK,CI}, and in particular has been of great
interest in the proposed `loop representation' (see, e.g. \cite{CR}) of quantum
gravity.    Dirac style methods are also commonly used in 
quantum cosmology, whether inspired by Einstein-Hilbert gravity
(e.g., \cite{Misner,MTW,RS,Wald} or string/M-theory  (e.g., \cite{BFM,CU}).

The Dirac proposal 
involves a large number of ambiguities.  As a result, one must make
certain choices in implementing the proposal, and those may
be roughly divided into two categories:  Category I has
to do with setting the stage for solving the constraints.
It includes, for example, both
the factor
ordering of the constraints and the choice of the space on which they should
act.  Category II has to do with implementing the rest of the scheme.
This includes the questions of in just what sense the constraints
should be `solved', how to make the solutions into a Hilbert space, and
how observables\footnote{Note
that in this paper all elements of the algebra of
observables will be called observables, and not just its self-adjoint 
elements.}
should then act on this space. 

 Choices in category
I are arguably shared with familiar non-relativistic
quantum mechanics, while category II choices are particular to Dirac's
scheme for constrained systems.  Thus, while one suspects that 
category I questions can be answered only in the context of particular
physical systems, category II questions could in principle have
quite general solutions.  The discovery of such solutions
could be of great help in the study of quantum gravity and/or cosmology.
Some efforts to address or at least to structure the discussion of
category II issues were made under the names of geometric quantization 
\cite{W,Tu,DE2,DE1}
(although this approach addresses certain category I issues as well),
Klein-Gordon methods \cite{Wald}, path integral methods 
\cite{HT,HH,HO}, coherent state quantization \cite{K2,K3},
$C^*$-algebra methods \cite{Grundling1,Grundling2}, 
algebraic quantization \cite{AAbook},
and refined algebraic quantization (see, for example \cite{ALMMT,GM} as
well as the closely related work of  \cite{AH,KL,QORD}).

Each of these uses a different strategy to implement the Dirac procedure, and, 
unfortunately, uses a different language to discuss the details. 
However, the Refined Algebraic Quantization (RAQ) scheme
has been shown to have a certain generality \cite{GM} 
and, in this sense, to include the other approaches.  This provides some
motivation for our study of RAQ in this work.  However, it should be
noted that RAQ is not at this point compatible with systems of constraints
involving structure {\it functions} instead of structure constants.
This means that it cannot even in principle be applied to full 3+1 gravity.
As will be discussed further in section \ref{Disc}, we hope that
our treatment of non-unimodular groups represents a step toward
removing this restriction.

Like the other approaches, RAQ codifies the Dirac scheme and formalizes
the choices that must be made for its completion.  RAQ first requires
that the system be quantized without imposing the gauge symmetry; that is, 
that a Hilbert space representation of the gauge-dependent operator
algebra be chosen.  In particular, the constraints are to be represented
as Hermitian operators.  This first Hilbert space is called the auxiliary
Hilbert space (${\cal H}_{\rm aux}$) to distinguish it from the 
physical Hilbert space to be constructed.  Choosing the constraint
operators and ${\cal H}_{\rm
aux}$ clearly falls under category I.

The next choice made by RAQ is that of a dense subspace $\Phi \subset
{\cal H}_{\rm aux}$ which is mapped into itself by the constraints.  It is
this choice which sets the final arena in which the constraints are
to be solved, as RAQ seeks solutions of the constraints in the
algebraic dual space $\Phi^*$ 
of linear functionals on $\Phi$.  To complete RAQ, one must
then find a map $\eta$ from $\Phi$ to $\Phi^*$ with certain special
properties.  This map is called the `rigging' map in vague analogy to the 
theory of rigged Hilbert spaces.  In RAQ, it simultaneously
solves the constraints (i.e., it's image contains only solutions)
and defines the physical Hilbert space ${\cal H}_{\rm phys}$.  The choice of
$\Phi$ and $\eta$  fall into Category II.  In the work
below, we will address the choice of the rigging map $\eta$.    One popular
idea for constructing $\eta$ is to use the group averaging procedure
(discussed below and in \cite{AH,ALMMT,BC}).
We find that, when the group averaging
integral converges sufficiently rapidly, the rigging map is unique (and given by
group averaging).  We hope to address the choice of $\Phi$ in future work.

The uniqueness of the group averaging map is our main result.  The surprising
feature is that this result holds subject only to the restriction that
the gauge group be a locally compact\footnote{Thus, the group must be
finite dimensional and our results do not apply directly to field theories.
Extensions to the infinite dimensional case may be possible using
path integrals but, at the level of rigor used here, the infinite dimensional
case is beyond the scope of this paper.} Lie group. In particular, it
applies to  arbitrary discrete groups (which we consider as zero-dimensional
Lie groups) and it includes the non-unimodular groups as well as non-amenable
groups (see \cite{Ki} for a definition)
and groups that are `wild' in the sense of \cite{Ki}
(non-type I, i.e., type-II and type-III groups in the terminology of
\cite{Mackey} and \cite{Dixmier}).

Arriving at a general result of this nature entails a significant
change of perspective from certain earlier works \cite{ALMMT,QORD,BC,spec}
on RAQ, which emphasized the connections of RAQ with spectral theory
and the decomposition of group representations as direct integrals of
Hilbert spaces. Roughly speaking, the idea there was to decompose the unitary 
representation of the gauge group on the auxiliary Hilbert space into a 
direct integral, and then select as the physical Hilbert space the integrand 
which carries the trivial representation.  This analogy is a powerful tool 
for dealing with Abelian groups (and one dimensional groups in particular), 
but leads one to expect certain (nonexistent) difficulties with non-amenable 
and wild groups. 

First of all, it is clear that mere `inspection' of a concretely given 
direct integral representation reveals the constituents only up to 
sets of measure zero. Hence a proper definition of `support' or 
`containment' of a representation needs to be given. This is achieved 
by the notion of {\it weak containment}, which, in brief, works as 
follows: Consider the space $C(G)$ of continuous complex-valued functions
on $G$ in the topology of uniform convergence on compact 
sets.\footnote{If $G$ is locally compact and Hausdorff, as it is here 
since we consider only finite-dimensional Lie groups, this topology is 
complete and equivalent to the compact-open topology (see \cite{Kelley}, 
chapter~7).} Matrix elements of representations of $G$ 
are now considered as elements in $C(G)$. Now, a representation $\rho$ 
is weakly contained in a family ${\cal S}$ of representations if the 
matrix elements of $\rho$ can be approximated by those of representations 
in ${\cal S}$. This defines a topology on the space of representations 
by defining the closure of a set as the set of weakly contained ones. 
For more information we refer to \cite{Fell} and also chapter 18.1 of 
\cite{Dixmier}. 

In the case of wild (non-type I) groups the direct integral decomposition 
into irreducibles of a representation is far more ambiguous over and above 
those trivial measure-theoretic ambiguities mentioned above.  In these 
cases it is therefore unclear how to interpret the appearance or absence 
of certain irreducible representations, and in particular the trivial 
representation, in the various direct integral decompositions.

The non-amenable case is characterized by the fact that the trivial 
one-dimensional representation is not weakly contained in the regular 
representation.\footnote{A group is amenable iff all its irreducible 
representations are weakly contained in the regular representation. 
It turns out that this is equivalent to the weak containment of just the 
trivial representation.}
Recall that in the regular representation the gauge 
group acts on the space $L^2(G,d_Lg)$ of square integrable functions 
with respect to
the left-invariant Haar measure $d_Lg$ on $G$. Now, non-amenable 
gauge groups suggest problems of the following kind: Consider a 
mechanical system with configuration space $G$ and left-translations 
of $G$ as gauge group. All momenta are then constrained to vanish and
the classical reduced phase-space consists of a single point.
Accordingly, one expects a single gauge invariant quantum state
to survive the quantization procedure. However, for a non-amenable 
group the trivial representation is not weakly contained, which suggests 
that one ends up with no states at all. Similar difficulties 
have been encountered for the action of the large diffeomorphisms on 
the state space of 2+1 gravity \cite{GL,CN} on $R \times T^2$. 

However, it turns out that refined algebraic quantization is
not in fact directly related to spectral analysis and weak containment.
Instead, it is more closely tied 
to another (coarser) topology on the space of group
representations.  This new topology is related to a new notion of
containment, which we call ``ultraweak containment,'' in the same way as 
the old topology was related to the notion of weak containment.
As will be discussed in section \ref{Disc} below, in order for group
averaging to produce a nontrivial result
it is necessary that a particular representation\footnote{The trivial
one, for unimodular groups.} be ultraweakly contained
in the representation of the gauge group on the auxiliary Hilbert space.
Under a natural restriction, the same is true of RAQ.
It would, of course, be of interest to show  (perhaps, along the lines of 
\cite{GM}), that ultraweak containment of the trivial representation is
a sufficient condition for the success of RAQ. 
However, no results of that sort are yet available.
It turns out that, even for non-amenable groups, the trivial representation
{\it is} always ultraweakly contained in the regular representation.
In addition, there is in general no direct integral
decomposition of representations in terms
of the irreducibles that they contain ultraweakly, so that there is no
distinction between wild and `tame' groups in this context.

Below, we briefly review Refined Algebraic Quantization and the method
of group averaging in section II.  
We then proceed in section III to show that, when group averaging
converges in a sufficiently strong sense, it gives the unique implementation
of Refined Algebraic Quantization (and therefore the unique implementation
of the Dirac procedure in the sense of \cite{GM}).  A key step in deriving
this result for the non-unimodular case is to realize
(following \cite{Tu} and \cite{DE1}) that, for such groups, the sense in 
which the constraints should be solved is somewhat different from what
one might 
naively expect.  However, as this represents a diversion from the main
thrust of the paper and as it will be of less interest to some readers, 
our discussion of this point is relegated to two appendices (A and B).
In the context of geometric quantization, 
this result was argued in \cite{Tu,DE1} by comparison with reduced
phase space methods and 
in \cite{DE2} by using the
fact that a gauge-system with non-unimodular gauge-group 
may always be suitably enlarged to a system with unimodular gauge group but 
isomorphic reduced phase space. This system can then be treated by known 
techniques including in particular the Dirac condition in its familiar form. 
But if projected to the original, non-unimodular, gauge system this then turns 
out to be equivalent to an alteration of the Dirac condition by an additional
term. References \cite{DE1} and \cite{DE2} also present some simple 
models of gauge-systems with non-unimodular gauge groups where the failure 
of the naive Dirac condition is studied explicitly. In Appendix~B we
summarize the unimodularisation technique just mentioned and also
discuss its group-averaging version.  Section \ref{Disc} 
contains a discussion of the results and introduces the topology of
ultraweak containment.

\section{Preliminaries}

In this section we briefly review both refined algebraic quantization
and group averaging.  More detailed treatments can be found in
\cite{ALMMT,GM,BC} and other references. In particular we refer the
reader to \cite{GM} for a thorough motivation of RAQ and some results
on its generality.

Refined Algebraic Quantization is a framework for the implementation
of the Dirac constraint quantization procedure which begins
by first considering an `unconstrained' quantum system in which even
gauge dependent quantum operators act on an auxiliary Hilbert space
${\cal H}_{\rm aux}$.  On this auxiliary space are defined self-adjoint
constraint operators $C_i$.  The commutator algebra of these quantum
constraints is assumed to close and form a Lie algebra\footnote{In particular,
this excludes systems such as General Relativity, which have structure functions
instead of structure constants.}.
Exponentiation of the operators will then yield a unitary representation of
the corresponding Lie group.  We will choose to formulate refined
algebraic quantization entirely in terms of this unitary representation
$U$ in order to avoid dealing with unbounded operators.

As with any version of the Dirac procedure, physical states 
$|\psi \rangle_{\rm
phys}$ must be
identified which in some sense solve the quantum constraints $C_i$.
Physically the same requirement is given\footnote{At least for
unimodular groups.  See Appendices A and B 
for a discussion of the non-unimodular
case.} by the statement that
the unitary operators $U(g)$ should act trivially on the physical 
states for any $g$ in the gauge group:
$ U(g) |\psi 
\rangle_{\rm phys} = |\psi \rangle_{\rm phys}$.  Now, as
the discrete spectrum of $U(g)$ need not contain one, 
the Hilbert space ${\cal H}_{\rm aux}$ will in general not contain any
such solutions.  However, by choosing some subspace $\Phi \subset {\cal
H}_{\rm aux}$ of `test states', we can seek solutions in the algebraic 
dual $\Phi^*$ of $\Phi$.  This means that we take $\Phi^*$ to be the 
space of all complex-valued linear functions on $\Phi$ and topologize this 
space by the topology of pointwise convergence, i.e., $f_n\rightarrow f$
in $\Phi^*$ iff $f_n(\phi)\rightarrow f(\phi)$ in the complex numbers 
for all $\phi\in\Phi$.
The space $\Phi$ should be chosen so that the operators $U(g)$ map $\Phi$ 
into itself. In this case, there is a well-defined dual action of $U(g)$
on $f \in \Phi^*$ given by  $U(g)f[\phi] = f[U(g^{-1})\phi]$ for all
$\phi \in \Phi$.  Solutions of the constraints are then elements
$f\in \Phi^*$ for which $U(g) f = f$ for all $g$. 

Of at least equal importance is a discussion of observables: gauge
invariant operators on ${\cal H}_{\rm aux}$ that will become operators
on the physical Hilbert space.  In RAQ, observables together with their
adjoints are required to include $\Phi$ in their domain and to map $\Phi$
to itself \cite{GM}. The star algebra of observables consists of all
such gauge invariant operators.
`Gauge invariance' of such an operator ${\cal O}$ then means that
${\cal O}$ commutes with the G-action on the domain $\Phi$:
${\cal O} U(g) |\phi \rangle=U(g) {\cal O} |\phi \rangle$ for all
$g\in G$, $\phi\in\Phi$.

Refined Algebraic Quantization then asks that a `rigging map' be chosen.
This is an anti-linear map from $\Phi$ into $\Phi^*$ satisfying the following
properties (here and in the rest of this work, an overline denotes 
complex-conjugation):

i) The image of $\eta$ solves the constraints. For unimodular $G$,
this means that for all $\phi_1, \phi_2 \in \Phi$ and all 
$g \in G$, $\eta(\phi_1)[U(g)\phi_2]=\eta(\phi_1)[\phi_2]$.
In the non-unimodular case, we will see that the right hand side needs 
to be multiplied by the square-root of $G$'s modular function 
(compare with equation (\ref{Diracmod})).

ii)  The map is real:
$\eta(\phi_1)[\phi_2] = \overline{\eta(\phi_2)[\phi_1]}$.

iii) The map is positive: $\eta(\phi_1)[\phi_1] \ge 0.$ 

iv) The map should commute with the observables or, in other words, 
intertwine the representations of the observables on $\Phi$ and $\Phi^*$:

\begin{equation}
{\cal O} ( \eta \phi ) = \eta ({\cal O} \phi).
\end{equation}
RAQ then constructs the physical Hilbert space by
introducing a Hermitian inner product on the image of $\eta$ and completing
this into a Hilbert space.  The inner product is just given
by the rigging map:

\begin{equation}
(\eta(\phi_1), \eta(\phi_2) )_{\rm phys} = \eta(\phi_1)[\phi_2].
\end{equation}
This is equivalent to defining a new inner product on $\Phi$, taking the quotient
by the zero norm vectors, and completing to a physical Hilbert space.

As stated in the introduction, the goal of this work is to study
the possible rigging maps.
A priori, it is not at all clear how large the space
of rigging maps might be.  See, for example, \cite {BC} for a discussion of
some of the
freedoms.  However, a natural idea is given by `group averaging' \cite{AH}.
This is the idea that one could construct such a rigging map by integrating
the operators $U(g)$ with respect to the Haar measure of the group.
Suppose for the moment that the group is unimodular so that the left
and right Haar measures agree.  We will call this (unique) Haar measure
$dg$.  In this case, the group averaging proposal is 
\begin{equation}
\label{ga1}
\eta |\phi \rangle := \langle \phi | \int dg \ U(g).
\end{equation}
Let us also, for the moment, ignore convergence issues and note that the 
expression  (\ref{ga1}) formally qualifies as a rigging map (except
perhaps for the positivity condition).  The invariance of the Haar
measure guarantees that any state in the image of the rigging map is
invariant under the action of $U(g_0)$ and so `solves the constraints.'
The reality and symmetry properties follow from the fact that
$dg = d(g^{-1})$.

When the group is non-unimodular, the left- and right- Haar measures
($d_L g$ and $d_R g$) do not agree, and are related by
$d_L g = [\Delta(g)]^{-1} d_R g$, where $\Delta(g)$ is the so-called
modular function, a homomorphism from the group $G$ to the positive real
numbers.  For finite dimensional Lie groups one has
$\Delta(g)=\det\{\Ad_g\}$, where $\Ad$ denotes the adjoint representation.
(See Appendix~A for details and and an explanation of our notation.)
Neither the right- nor the left- Haar measure is invariant under the map
$g \rightarrow g^{-1}$. However, there is another measure, halfway between
the left- and right- Haar measure, which is invariant under
$g \rightarrow g^{-1}$.
We will call this measure $\d g$, the symmetric measure.  It is given
by $\d g =  [\Delta(g)]^{1/2} d_L g =  [\Delta(g)]^{-1/2} d_R g$
(see Appendix~A).

We will work below
with the entire class of measures
\begin{equation}
d_n g = \Delta^{n/2}(g) d_0g, \ {\rm for } \ n \in {\bf Z}.
\end{equation}
In particular, $d_Lg = d_{-1} g$ and $d_R g = d_1 g$.
Note that these measures satisfy the following properties:
\begin{eqnarray}
\label{measure rels}
d_n g^{-1} = d_{-n} g = \Delta^{-n}(g) d_n g \cr 
d_n (ag) = \Delta^{{n+1} \over 2} (a) d_n g \cr
d_n (ga) = \Delta^{{n-1} \over 2} (a) d_n g.
\end{eqnarray}

Thus, the following expression, replacing (\ref{ga1}),
\begin{equation}
\label{GA}
\eta |\phi \rangle := \langle \phi | \int \d g \  U(g) 
\end{equation}
has the right properties for $\eta$ to be real in the sense of RAQ.
However, this expression is not invariant under multiplication on the
right by $U(g_0)$.  Instead, $(\eta | \phi \rangle) U(g_0)
= (\eta | \phi \rangle) \Delta^{1/2}(g_0)$.  As argued in
Appendix~B (based on \cite{DE1,DE2,Tu}), this is in fact the correct result in
the non-unimodular case.  If we consider RAQ in terms of defining a new
inner product on $\Phi$, then the group averaging inner product is just
\begin{equation}
\label{GAM}
\langle \phi_1 | \phi_2 \rangle_{ga} = \int \d g \langle \phi_2  | U(g) | \phi_1
\rangle.
\end{equation}

In order for the group averaging procedure to be well defined, the integral
(\ref{GAM}) must converge absolutely for all $\phi_1$ and $\phi_2$ in $\Phi$.  
However, to complete our uniqueness proof, we will need to impose the 
somewhat stronger condition that the integrals

\begin{equation}
\int d_n g \langle \phi_2  | U(g) | \phi_1
\rangle.
\end{equation}
converge absolutely for all $n \in {\bf Z}$.
In this case the function
$\langle \phi_2  | U(g) | \phi_1\rangle$ defines an element of $L^1(G, d_n g)$
for all $n$ and we will refer to $\Phi$ as an $L^1$ space. Similarly, a
state $\phi$ will be called $L^1$ when
$\langle \phi  | U(g) | \phi\rangle$ is in $L^1(G,d_n g)$ for all $n$.
In this case, 
we denote the function 
$\langle \phi  | U(g) | \phi \rangle $  by $U_\phi(g)$.

\section{A Unique Rigging Map}

This section contains the proof of our uniqueness theorem.  A key
feature of our derivation will be the use of a certain group algebra
for $G$, which we introduce and study in subsection A below.  The main
result then follows in subsection B.

\subsection{The group algebra}
\label{ga}

We now construct a group algebra ${\cal A}_G$ based on the
functions on $G$ that are
$L^1$ with respect to {\it all} of the measures $d_n g$ for $n \in {\bf Z}$.
We will denote the 
$L^1(G, d_n g)$ norm of $f$ by
$||f||_n$.   The group
algebra multiplication of two functions $f_1, f_2 \in {\cal A}_G$ will
be defined to be convolution with respect to the right Haar measure:
\begin{equation}
\label{mult}
f_1 \cdot f_2 (g) =  \int_G (d_R g_1) \  f_1(gg_1^{-1}) f_2(g_1).
\end{equation}

To see that $f_1 \cdot f_2$ is in $L^1(G, d_n g)$, 
note that
\begin{equation}
|f_1 \cdot f_2 (g) |  \leq  \int_G d_R g_1  |f_1(g g_1^{-1})|
 \ |f_2 (g_1)|.
\end{equation}
Thus, it is sufficient to check the result for positive $f_1$ and $f_2$.
Consider such positive real $f_1, f_2$ and
consider some compact subset $K \subset G$.  We have
\begin{equation}
\int_K d_n g \ (f_1 \cdot f_2) =  \int_K d_n g \int_G d_R g_1 \  
f_1(gg_1^{-1}) f_2(g_1) = \int_G d_n g_1 
\left (\int_{Kg_1^{-1}} d_n g_2 \ f_1(g_2)
\right) f_2(g_1).
\end{equation}
Since the integral in parentheses converges absolutely to $||f_1||_n$
in the limit as $K \rightarrow G$, it follows that the entire expression
converges to $||f_1||_n \ ||f_2||_n$.  Thus, for any $f_1, f_2 \in 
{\cal A}_G$, we have $||f_1 \cdot f_2 ||_n \leq ||f_1||_n \ ||f_2||_n$, 
with equality if and only if $f_1$ and $f_2$ are real and positive.
It follows that the operation $f_1 \cdot f_2$ defines
a product on ${\cal A}_G$.

The group algebra ${\cal A}_G$ is in fact a star algebra.  The
star operation is defined by
\begin{equation}
f^\star(g) = \overline{f(g^{-1})}.
\end{equation}
{}From (\ref{measure rels}), we have that $||f||_n = ||f||_{-n}$, 
so ${\cal A}_G$ is closed under the action of $\star$.  From (\ref{mult})
it follows that $(f_1 \cdot f_2)^\star = f_2^\star \cdot f_1^\star$.
Another (linear) involution on ${\cal A}_G$ of which we  will make use
is $*$, given by
\begin{equation}
\label{*}
f^*(g) = f(g^{-1}) \Delta^{- \frac{1}{2}}(g).
\end{equation}

Now, the important part of this construction is that, to each 
$f \in {\cal A}_G$, we may associate an operator $\hat{f}$ on 
${\cal H}_{\rm aux}$.  This operator
is given by
\begin{equation}
\hat f = \int_G \d g \  f(g) U(g).
\end{equation}
Note that the operator norm of $\hat{f}$ is bounded by $||f||_0$, so
that $\hat{f}$ is defined on all of ${\cal H}_{\rm aux}$. As may be
readily verified, the map $f \mapsto \hat{f}$ 
satisfies $\widehat{f^\star}=(\hat{f})^\dagger$ and 
$\widehat{f_1 \cdot f_2}=\hat f_1\hat f_2$.  Hence it defines a 
$\star$-homomorphism from ${\cal A}_G$ to $B({\cal H}_{\rm aux})$,
the bounded operators on ${\cal H}_{\rm aux}$. 
Thus, the group algebra ${\cal A}_G$ acts on ${\cal H}_{\rm aux}$
via its image ${\hat{\cal A}}_G\subset B({\cal H}_{\rm aux})$. 

An important property of this action is that it
preserves the $L^1$ property of
a state and of a subspace.  
To see this, we first prove the result:

\proclaim Lemma 1.  If $\phi$ is an $L^1$ state, then  $\langle
\phi | \hat{f}_1^\dagger U(g) \hat{f}_2 | \phi \rangle$ is in
$L^1(G, d_n g)$.

\beginproof
For a compact subset $K \subset G$, consider the expression 
\begin{equation}
\int_K d_n g \ | \langle  \phi | \hat{f}_1^\dagger U(g) 
\hat{f}_2 |\phi \rangle|
= \int_K d_n g \int_{G\times G} d_0 g_1 \ d_0 g_2 \  | \overline{f}_1 
(g_1) f_2(g_2) U_\phi(g_1^{-1} g g_2) |.
\end{equation} 
Setting $t = g_1^{-1} g g_2$ and replacing $g$ by $g = g_1 t g_2^{-1}$
leads to the expression
\begin{equation}
\int_{G\times G} \d g_1 \ \d g_2  \int_{g_1^{-1} K g_2} d_n t \
|\overline{f}_1 (g_1) f_2(g_2)
U_\phi(t) \Delta^{{1+n} \over 2} (g_1) \Delta^{{1-n} \over 2}(g_2)|.
\end{equation}
In the limit as $K \rightarrow G$ (and hence $g_1^{-1}Kg_2\rightarrow G$), 
the $t$-integral converges absolutely and the factors of $\Delta^{1/2}$ 
combine with the symmetric measures to make $d_{n+1}g_1$ and $d_{1-n} g_2$. 
Thus, $\langle \phi | \hat{f}_1^\dagger U(g) \hat{f}_2 | \phi \rangle$ is in 
$L^1(G, d_n g)$ and $\int_G d_n g | \langle 
\phi | \hat{f}_1^\dagger U(g) \hat{f}_2 
| \phi 
\rangle| = ||f_1||_{1+n} \ ||f_2||_{1-n} \ ||U_\phi ||_{n}$.  In particular, if
$\phi$ is an $L^1$ state then so is $\hat{f} \phi$.
Thus, we arrive at the conclusion that the space of $L^1$ states
carries a $\star$-representation of the group algebra ${\cal A}_G$.
\endproof

Two immediate extensions of the above Lemma should be noted.  The first
is that the result continues to hold if either $\hat{f}_1$ or $\hat{f}_2$
is replaced by $U(g_0)$ for any $g_0 \in G$.   Thus, we may extend
Lemma 1 to include any $\hat f_1, \hat f_2$ in the algebra $\hat {\cal A}_G$
generated by $\hat f$ for $f \in {\cal A}_G$ and $U(g_0)$ for $g_0 \in G$.
To state the second, let us
say that $\phi_1$ and $\phi_2$ are $L^1$ with respect to each
other when $\langle \phi_1 | U(g) | \phi_2 \rangle$ lies in $L^1(G, d_n g)$. 
It then follows that,  
if $\phi_1$ and $\phi_2$ are $L^1$ with respect to each other, 
then so are $\hat{f}_1 \phi_1$ and $\hat{f}_2 \phi_2$ for any
$\hat f_1, \hat f_2 \in \hat {\cal A}_G$.  To see this, simply  let
$M(\phi;g) = 
\langle \phi | \hat{f}^\dagger_1 U(g) \hat{f}_2 | \phi\rangle$ and
note the usual polarization identity:

\begin{equation}
\langle \phi_1 | \hat{f}^\dagger_1 U(g) \hat{f}_2 | \phi_2 \rangle
= [M(\phi_1 + \phi_2;g) - M(\phi_1 - \phi_2;g) - i M(\phi_1 +i
\phi_2;g) + i M(\phi_1 - i \phi_2;g)]/4 .
\end{equation}
Taken together, these results give the following:

\proclaim Lemma 2.  If $\Phi$ is an $L^1$ subspace of ${\cal H}_{\rm aux}$,
then  so is any space generated by the action of $\hat {\cal A}_G$ on $\Phi$.

\subsection{The Main Result}

We will now show that, given an $L^1$ subspace $\Phi$ that is
invariant under the action of $\hat {\cal A}_G$ and on which the group
averaging inner product (\ref{GAM}) is not identically zero, any
rigging map is some constant times the group averaging map.
Note that, given any $L^1$ space $\Phi_0$, the results of subsection
\ref{ga}
show that an invariant $L^1$ space can be constructed by taking the 
closure of $\Phi_0$ under the action of $\hat {\cal A}_G$.

It is important to remark
that we do not know whether the group averaging map is in general
positive, and that we will {\it not} need to assume positivity of the
group averaging map in this section.  In particular, this means that if in
some case the group averaging map fails to be positive semi-definite, then there
{\it cannot} exist a positive rigging map.  No cases of this sort are yet
known, but neither are they ruled out by general theorems known to the authors.

The most restrictive property of the rigging map is that it must intertwine
representations of the observable algebra ${\cal A}_{\rm obs}$ acting
on $\Phi$ and $\Phi^*$.  Our uniqueness theorem will follow by 
making use of a certain class of observables.  These observables
are of the sort constructed in \cite{QORD} by averaging gauge-dependent
operators over the gauge group.  To this end, we will need the following
Lemma:

\proclaim Lemma 3.  Let $\phi_1, \phi_2$ be states in the invariant
$L^1$ space $\Phi$.  
The expression
\begin{equation}
\label{gaos}
{\cal O}_{\phi_1 \phi_2} = \int d_L g U(g)
|\phi_1 \rangle
\langle \phi_2| U(g^{-1}) 
\end{equation}
defines an observable.

\beginproof
Recall that, in the context of RAQ, an observable is an operator
whose domain includes our chosen dense subspace $\Phi$, maps 
$\Phi$ to itself, and which commutes with the action of the gauge group
on the domain $\Phi$.  Let us therefore consider the action of 
${\cal O}_{\phi_1,\phi_2}$
on some state $\phi_0 \in \Phi$.  We introduce the notation
$U_{\phi_2 \phi_0}(g) = \langle \phi_2 | U(g) | \phi_0 \rangle$.  Note that
since $\phi_2$ and $\phi_0$ 
are in $\Phi$, we know that $U_{\phi_2 \phi_0}$
is $L^1$ with respect to each of the measures $d_ng$, and so defines
an element of ${\cal A}_G$.  Recalling the definition of the $*$ involution
from (\ref{*}), we may write
\begin{equation}
\label{12act}
{\cal O}_{\phi_1, \phi_2} |\phi_0 \rangle
= \int d_L g U_{\phi_2 \phi_0} (g^{-1}) U(g) 
|\phi_1  \rangle = \widehat{ {U_{\phi_2 \phi_0}^*}} | \phi_1
\rangle.
\end{equation}
As an element of $\hat{\cal A}_G$, the operator 
$\widehat{ {U_{\phi_2 \phi_0}^*}}$ is bounded, so that the
integral in (\ref{12act}) converges in the Hilbert space norm. 
The domain of ${\cal O}_{\phi_1 \phi_2}$ thus
contains all of $\Phi$.   Furthermore, since $\Phi$ is invariant
under the action of $\hat {\cal A}_G$, equation (\ref{12act}) shows
that ${\cal O}_{\phi_1 \phi_2}$ maps $\Phi$ into itself.
Since the measure in (\ref{gaos}) is invariant under
left translations, it follows immediately that ${\cal O}_{\phi_1 \phi_2}U(g)
|\phi_0 \rangle =
U(g) {\cal O}_{\phi_1 \phi_2} |\phi_0 \rangle$ for any $\phi_0 \in \Phi$, 
and that ${\cal O}$ is an observable. 
\endproof

\noindent Note that the operators (\ref{12act}) in fact define a complete
set of operators on the physical Hilbert space in the sense that they
generate the entire algebra of bounded operators.

We are now ready to prove our main result.

\proclaim Theorem.  Suppose that $\Phi$ is an $L^1$ subspace of
${\cal H}_{\rm aux}$
which is invariant under the action of ${\hat{\cal A}}_G$ and on which
the group averaging bilinear form (\ref{GAM}) is not identically zero.
Then the rigging map is unique up to an overall scale.
Furthermore, it is given by the group averaging expression (\ref{GA}).

\beginproof
We begin by choosing any two states $\phi_1$ and $\phi_2$ in $\Phi$ and
supposing that $\eta$ is a rigging map.  It suffices to consider the case
where $\eta$ is not the zero map.  Associated with these states are the
observables ${\cal O}_{\phi_1 \phi_2}$ and ${\cal O}_{\phi_2 \phi_1}$
defined above. As observables, ${\cal O}_{\phi_1\phi_2}$ and 
${\cal O}_{\phi_2\phi_1}$ act in the physical Hilbert space and 
$\eta$ must intertwine this action with the action of these operators on 
$\Phi$. 
We now consider their action on the physical states 
$\eta\phi_1$ and $\eta \phi_2$. For this we note that 
$\eta\hat f=\left(\int_G d_1 g f(g) \right)\eta$ for any $f\in{\cal A}_G$.
Then, for $i,j$ ranging over $1,2$, we have
\begin{eqnarray}
\label{Sact}
{\cal O}_{\phi_i \phi_j} \eta \phi_j & = &
\eta {\cal O}_{\phi_i \phi_j} \phi_j
= \eta \widehat{{U^*}}_{\phi_j 
\phi_j}  \phi_i \cr
&=&  
\left[ \int_G d_0 g \langle \phi_j | U(g) | \phi_j \rangle \right]
\eta \phi_i. 
\end{eqnarray}
Finally, recall (e.g. \cite{ALMMT,GM}) that any rigging 
map preserves $*$-relations\footnote{Proof: 
${\cal O}\eta(\psi_2)[\psi_1]
  =\eta(\psi_2)[{\cal O}^{\star}\psi_1]$ 
   (by definition of dual action)
$ =\overline{\eta({\cal O}^{\star}\psi_1)[\psi_2]}$ 
   (by reality)
$ =\overline{{\cal O}^{\star}\eta(\psi_1)[\psi_2]}$
   (by intertwining property).}
in the sense that, since $\langle \psi_2 | {\cal O}_{\phi_1 \phi_2} |
\psi_1 \rangle = \overline{ \langle \psi_1 | {\cal O}_{\phi_2 \phi_1}
| \psi_2 \rangle}$ for all $\psi_1, \psi_2 \in \Phi$, we also have
\begin{equation}
{\cal O}_{\phi_1 \phi_2}\eta(\psi_2)[\psi_1] = \overline
{{\cal O}_{\phi_2\phi_1}\eta(\psi_1)[\psi_2]}.
\end{equation}
Taking $\psi_1 = \phi_1$, $\psi_2 = \phi_2$, we apply
(\ref{Sact}) with $i=1,j=2$ to the left hand side and with 
$i=2,j=1$ to the right hand side. Taking into account the hermiticity
of $\int_G d_0 g \langle \phi_i | U(g) | \phi_i \rangle$, and
the hermiticity of the physical inner product, we arrive at 
\begin{equation}
\label{result}
\left( \int_G d_0 g \langle \phi_1 | U(g) | \phi_1 \rangle \right)
\eta(\phi_2)[\phi_2] = 
\left( \int_G d_0 g \langle \phi_2 | U(g) | \phi_2 \rangle \right)
\eta(\phi_1)[\phi_1]
\end{equation}
which must hold for all $\phi_1, \phi_2 \in \Phi$.

Now, both Hermitian forms, the one defined by group averaging and 
the one defined by $\eta$, are non-zero. Suppose $\phi_1$ is in the 
kernel of the first form. Choosing $\phi_2$ in the complement of this 
kernel, (\ref{result}) implies that $\phi_1$ is also in the kernel 
of the second form. Reversing the r\^oles of the two forms in this 
argument shows that their kernels coincide. Equation (\ref{result}) 
also implies that in the complement of this kernel the ratio 
$k := { {\eta(\phi_1)[\phi_1]}
\over{\left(\int_G d_0 g\langle\phi_1 | U(g) | \phi_1\rangle\right)}}$
is a constant, independent of $\phi_1$.  Furthermore, the rigging
map $\eta$ is just this constant $k$ times the group averaging map
defined by the symmetric inner product $d_0 g$.  
\endproof

\section{Discussion}
\label{Disc}

In this work we have shown that, when group averaging converges
sufficiently quickly (and gives a nontrivial result), 
it defines the unique rigging map and thus the
unique physical inner product.  Convergence of group averaging is not
uncommon for finite dimensional gauge groups.  A particular case
of interest is the group ${\bf R}$, which has a single generator.
This case includes the quantization of homogeneous cosmological
models, in which the single generator is the Hamiltonian constraint.
The convergence of group averaging is most easily analyzed by considering
a complete set of (generalized) eigenstates of the self-adjoint constraint.
So long as such states can be parameterized by points on a smooth manifold, 
it is usually not difficult to construct an $L^1$ domain $\Phi$ from
smooth functions on that manifold and to
see that group averaging defines a positive rigging map.  
A dense $L^1$ space is also easy to find in the (say left-) regular 
representation of any group $G$, where $G$ acts unitarily on $L^2(G,d_Lg)$
via $(U(h)\phi)(g):=\phi(h^{-1}g)$. One need only take the space 
$\Phi$ to be the closure under the action of $\hat {\cal A}_G$ on the 
space of functions of compact support on $G$, and all integrals are 
guaranteed to converge. Moreover, for the regular (and hence any sub-) 
representation the group averaging map is positive. This follows from 
\begin{equation}
\label{positivity}
\int_Gd_0h\langle\phi,U(h)\phi\rangle_{L^2}=
\int_{G\times G}d_0h\,d_Lg\ \overline{\phi(g)}\, \phi(h^{-1}g)=
\left\vert\int_Gd_0g\, \phi(g)\right\vert^2\geq 0.
\end{equation}
Further remarks on positivity may be found in \cite{Rief}.

Our result is quite strong, as it applies to an arbitrary {\it locally} 
compact group, even if non-unimodular, non-amenable, or wild.    
The discussion of non-unimodular groups is especially interesting and 
may hold important insights for future work. Following \cite{Tu,DE2,DE1}, 
our physical states `solve' the equations
\begin{equation}
\label{Diracmod}
U(g) | \phi \rangle_{\rm phys} = \Delta^{1/2} | \phi \rangle_{\rm phys},
\end{equation}
where $U(g)$ is a unitary representation of the gauge group $G$.
However, this is equivalent to introducing the non-unitary representation
$\rho(g) = \Delta^{- \frac{1}{2}}(g)U(g)$ and taking physical states
to be invariant under this action.  This may represent
the first step in moving away from self-adjoint constraints
and unitary group actions which, as argued in \cite{Karel}, may be necessary
for a treatment of gravity-like systems where  the presence of
structure functions means that the constraints do not form a Lie algebra.

Even staying within the 
context of Lie groups and Lie algebras, there are clearly a number of 
important issues remaining, such as understanding what happens when
group averaging does not converge sufficiently rapidly (or at all!) or
when the group averaging inner product is zero.
In fact, based on
our result, one expects this to  be a common occurrence.
To see this, suppose that there is some superselection rule
(in the sense of \cite{ALMMT}) on ${\cal H}_{\rm aux}$.  By this we
mean that the set of observables defined as above for some choice of $\Phi$
leaves invariant some nontrivial decomposition ${\cal H}_1
\oplus {\cal H}_2$ of ${\cal H}_{\rm aux}$.  Let $\eta_1, \eta_2$ be the
restrictions of any rigging map $\eta$ to ${\cal H}_1$ and ${\cal H}_2$
respectively.  Then, it is clear that any map of the form $a_1 \eta_1
\oplus a_2 \eta_2$ is also a rigging map when $a_1, a_2$ are real and positive. 
So long as both $\eta_1$ and $\eta_2$ are nontrivial, this will
give a real ambiguity in the choice of rigging map.  Thus, by our result
above, group averaging {\it cannot} converge rapidly on any 
subspace in the common domain of $\eta_1$ and $\eta_2$ where both
rigging maps act nontrivially.
In such cases, one might hope that some procedure related to group
averaging can be applied (perhaps along the lines of the ``renormalization'' of
group averaging used in \cite{ALMMT}).  Some exploration in this direction
will appear in \cite{GoMa}.  

As remarked above, our uniqueness theorem holds
regardless of non-amenability or the possible `wild' character of 
a group.  In the case of a non-amenable group, the regular representation
does not weakly contain the trivial representation, so the trivial 
representation does not appear in a direct integral decomposition.
Nonetheless, as we have seen,  group averaging
{\it does} succeed in producing a trivial representation from the regular
representation. 
Thus, it is clear that refined algebraic quantization in general (and
group averaging in particular) is not tied to the concept of weak
containment.
However, if we are to move beyond our work here and study in detail
the space of rigging maps in the case where group averaging does not
converge, it would be useful to find a similar concept which might be
used in place of weak containment.  Our suggestion is the notion of
`ultraweak containment,' which we now introduce.

Ultraweak containment is a topology on the space of {\it all} representations
of a group $G$.  It is explicitly our intention to
include non-unitary representations (so that we may discuss the
representation $\Delta^{1/2}$ associated with non-unimodular groups).
We specify
this topology by stating when a representation $\rho_0$ lies in the
closure of a set $R$ of representations. 

\proclaim Definition.  A representation $\rho$ of a group $G$
on a Hilbert space ${\cal H}$ lies in the ultraweak
closure of a set $R$ of Hilbert space representations of $G$
when, for every pair of states $\phi_1, \phi_2 \in {\cal H}$ there exists
a sequence of triples $\{(\rho_n, \phi_1^n, \phi_2^n) \}$, of representations
$\rho_n \in R$ and states $\phi_1^n$ and $\phi_2^n$ which lie in the
Hilbert space carrying $\rho_n$, which satisfy
\begin{equation}
\label{42}
\lim_{n \rightarrow \infty}
\langle \phi_1^n | \rho_n(g) | \phi_2^n \rangle = 
\langle \phi_1 | \rho(g) | \phi_2 \rangle.
\end{equation}
as functionals on $C_0^\infty(G)$; i.e., that 
\begin{equation}
\label{uw}
\lim_{n \rightarrow \infty} \int_G d_0g \ f(g)
\langle \phi_1^n | \rho_n(g) | \phi_2^n \rangle =  \int d_0g \ f(g)
\langle \phi_1 | \rho(g) | \phi_2 \rangle,
\end{equation}
for all smooth functions $f$ of compact support on $G$. Note that
this definition differs from weak containment just in the choice
of topology in which limit (\ref{42})
is to converge: Instead of the 
compact open topology on $C(G)$ we now use the topology of
$C_0^{\infty}$'s dual.

\noindent
Note that the sequence is not allowed to depend on the group element $g$ (or,
on the test function $f$),
but that as the sequence depends on the {\it pair} $\phi_1, \phi_2$,
taking $\phi_1 = \phi_2$ does not require taking $\phi_1^n = \phi_2^n$.

The following result shows the relevance of this topology to group
averaging, and also illustrates the fact that this topology is
distinctly non-Hausdorff.

\proclaim Proposition.
Consider any unitary representation $\rho$ of $G$ on a Hilbert
space ${\cal H}$ having an $L^1$ subspace $\Phi$ on which group
averaging yields a nontrivial physical Hilbert space.  The representation
$\Delta^{1/2}$ lies in the ultraweak closure of the single element
set $\{ \rho \}$.

\beginproof
To see this, simply choose any state $\phi_0 \in \Phi$ and a sequence
of compact sets $K_n \subset G$ which expand to all of $G$.  We
let $\phi_2^n = k\phi_0$, where $k$ is some number to be determined,
and $\phi_1^n = \int_{K_n} d_0g \ \rho(g) |\phi_0 \rangle.$  It follows
that $\lim_{n \rightarrow \infty} \int_G f(g) \langle \phi_1^n | \rho(g)
| \phi_2^n \rangle d_0g
= k\,\lim_{n \rightarrow \infty}  \int_G d_0g\,f(g) \ \int_{K_n} d_0 g_1
\langle \phi_0 | \rho(g_1) \rho(g) | \phi_0 \rangle = \left[ \int_G d_0g 
\Delta^{1/2}(g) f(g) \right]\, k\,
\left(\int_G d_0g_2 \langle \phi_0 | \rho(g_2)| \phi_0 \rangle \right)$.
Choosing $k^{-1}=\int_G d_0g_2\langle\phi_0 | \rho(g_2) | \phi_0\rangle$
shows the result.
\endproof

A similar result holds whenever RAQ can be implemented on a space $\Phi$
which is dense in its dual $\Phi^*$.  The restriction of RAQ to
this case may be natural on the grounds that $\Phi^*$ should provide
a `completion' of $\Phi$.  With this understanding, ultraweak
containment of the $\Delta^{1/2}$ representation is essential to the
success of RAQ.

\acknowledgements
D.G. was supported by the Swiss National Science Foundation,
the Tomalla Foundation and the Albert Einstein Institute.
D.M. was supported in part by National Science Foundation
grants PHY94-07194 and PHY97-22362, by
funds from Syracuse University, and by the
Albert Einstein Institute.

\appendix

\section{Measures on Lie Groups}

Let $G$ be a finite dimensional, connected  Lie group.
The maps of left and right multiplication with $g\in G$ are
denoted by $L_g$ and $R_g$ respectively; push-forwards and
pull-backs of maps are denoted by a lower and upper $*$.
By $\Ad$ and $\Ad^*$ we denote the adjoint representation of
$G$ on its Lie algebra $\g$ and its transpose, the
anti-representation on the dual $\g^*$ of the Lie algebra.
Note that $g\mapsto \Ad^*_{g^{-1}}$ is a proper representation 
of $G$ on $\g^*$.

If $n=\hbox{dim}(G)$, let $\mu$ be an $n$-form over the
tangent-space at the identity $e\in G$. We extend it to
left- and right-invariant $n$-forms over $G$ by defining
$\mu_L(g):=L_{g^{-1}}^*\mu$ and $\mu_R(g):=R_{g^{-1}}^*\mu$
respectively. $G$ is unimodular iff $\mu_L=\mu_R$.
But we have $R^*_h\mu_L(g)=R^*_hL^*_{(gh)^{-1}}\mu
=L^*_{g^{-1}}\Ad_{h^{-1}}^*\mu=\det\{\Ad_{h^{-1}}\}\mu_L(g)$,
so that for Lie groups unimodularity of $G$ is equivalent to
(hence its name) $\Delta(g):=\det\{\Ad_g\}=1\,\forall g\in G$.
Note that $\Delta:G\rightarrow R_+$ is a homomorphism into the
multiplicative group of the positive real numbers. Let
$I:g\mapsto g^{-1}$ denote the inversion map, then clearly
$L_h\circ I=I\circ R_{h^{-1}}$ and $I^*_e\mu=(-1)^n\mu$, implying
$I^*\mu_L=(-1)^n\mu_R$. We define the `inversion-symmetric'
$n$-form $\mu_0:=\Delta^{1\over 2}\mu_L=\Delta^{-{1\over 2}}\mu_R$
which satisfies  $I^*\mu_0=(-1)^n\mu_0$.

Suppose some $\mu$ has been fixed, we then write $d_Lg:=\mu_L(g)$,
$d_Rg:=\mu_R(g)$ and $\d g=\mu_0(g)$. From now on we agree that the
integration variable is always $g$. We then write $(L_h^*{\t \mu})(g)
=:{\t d}(hg)$, $(R_h^*{\t \mu})(g)=:{\t d}(gh)$ and
$(I^*{\t \mu})(g)={\t d}(g^{-1})$, where $\t\mu$ stands for
$\mu_L,\mu_R,\mu_0$ and $\t d$ for $d_L,d_R,\d$ respectively.
{}From the above one  easily shows that:
\begin{equation}
\label{measures}
\begin{array}{ll}
 d_L(gh)=\Delta^{-1}(h)\,d_Lg,
&\qquad d_L(g^{-1})=(-1)^n\Delta(g)\,d_Lg,             \\
 d_R(hg)=\Delta(h)\,d_Rg,
&\qquad d_R(g^{-1})=(-1)^n\Delta^{-1}(g)\,d_Rg,        \\
 \d (hg)=\Delta^{1\over2}(h)\,\d g,
&\qquad\d (gh)=\Delta^{-{1\over2}}(h)\,\d g.           \\
\end{array}
\end{equation}
Note that the factor $(-1)^n$, which results from the orientation
reversing nature of $I$ in case of odd-dimensional $G$, does not
appear in the integrals. Hence we also suppressed it in the
transformation formulae for the measures in the main text.

\section{Non-Unimodular Gauge Groups and their Unimodularisation}

In this appendix we explain in some detail the process of
unimodularisation, which allows one to reduce the cases of
non-unimodular gauge groups to the unimodular ones. Our discussion
is divided into three parts: in the first we explain the classical
aspect and in the second its realization in geometric quantization.
Here we basically follow ref. \cite{DE1}. In the third part we
show that this procedure fits naturally into the framework of group
averaging. In the following we assume familiarity with the basic
concepts of symplectic mechanics, including the notion of momentum maps
(see Chapt.~4 of \cite{AM} or Appendix~5 of \cite{Ar}), and 
differential geometry.

\subsection{Unimodularisation Classically}

We consider a finite dimensional first-class
gauge system $(M,\omega,G,\Phi, J)$, consisting of an even
dimensional manifold $M$, a symplectic structure $\omega$, a
gauge group $G$, a symplectic left action\footnote{Here we
follow standard notational conventions of geometric quantization.
This left-action $\Phi$ should not be confused with the dense subspace
$\Phi$ of RAQ (which will make no appearance in this appendix).}
$\Phi:G\times M\rightarrow M$ and an $\g^* $-valued,
$\Ad^*$-equivariant (i.e. $J\circ\Phi_g=\Ad^*_{g^{-1}}\circ J$) 
momentum map $J:M\rightarrow\g^*$. That the gauge system is 
first-class means that the constraint submanifold, given by 
$J^{-1}(0)\subset M$, is coisotropic, i.e. 
$T^{\perp}(J^{-1}(0))\subset T(J^{-1}(0))$, where $\perp$ refers to 
the $\omega$-orthogonal complement. In order for the reduced phase 
space to become a manifold we tacitly assume here $\Phi$ to be proper 
and free\footnote{Freeness is not really necessary. It may be 
relaxed to the condition that the stabilizer subgroups of the 
action are all conjugate. See exercise 4.1M. on p. 276 of \cite{AM}.}. 
The constraint set still supports a left action of $G$, and the 
reduced phase-space is the orbit space of  this action:
$M_{\rm red}:=G\backslash J^{-1}(0)$, with its symplectic structure
$\omega_{\rm red}$ induced by $\omega$. Now the point is that if 
$G$ is not unimodular,
we can always replace $(M,\omega, G, \Phi, J)$ 
in a canonical fashion by some `larger' 
$(\t M, \t\omega, \t G, \t\Phi,\t J)$, such that (i) $\t G$ is 
unimodular and (ii) $({\t M}_{\rm red},{\t\omega}_{\rm red})$ is
symplectomorphic to $(M_{\rm red},\omega_{\rm red})$. This process, which
may be understood as a special case of symplectic induction (\cite{GS}), 
is called {\it unimodularisation}. In many aspects it closely 
resembles the BRST-method, albeit in a purely bosonic setting.  

More concretely, unimodularisation starts with the choice $\t G=T^*G$. 
We globally
trivialize $T^*G$ by left translations, i.e. we map $\alpha_g\in T_gG$ to
$(g,L_g^*\alpha_g)\in G\times\g^*$, and identify (as manifolds)
$T^*G$ with $G\times\g^*$. The group structure is that of a semi-direct
product, with $G$ acting on $\g^*$ via $(\Ad^*)^{-1}$, so that we can
write $\t G=G\ltimes\g^*$ and have 
$(g,\alpha)(h,\beta)=(gh,\alpha+\Ad^*_{g^{-1}}(\beta))$ and
$(g,\alpha)^{-1}=(g^{-1},-\Ad^*_g(\alpha))$.  Using these formulae
it is straightforward to show that the matrix of the adjoint
action of $(g,\alpha)\in\t G$ on $\t\g\cong\g\oplus\g^*$ has the
block from
\begin{equation}
\label{App1}
\pmatrix{\Ad_g & 0             \cr
             * & \Ad^*_{g^{-1}}\cr}.
\end{equation}
Hence it has unit determinant so that $\t G$ is unimodular.

Further, one chooses  $\t M=M\times\g^*\oplus\g$ with $\g^*\oplus\g$
understood as cotangent bundle $T^*\g^*$, carrying its canonical
symplectic 
structure\footnote{In the standard Darboux-coordinates $(q,p)$
for (base space,fibre), this is just the familiar  
$\hat\omega=dp\wedge dq=d\sigma$ with $\sigma=pdq$.}
$\hat{\omega}=d\sigma$, where
$\sigma_{(\alpha,a)}(\beta,b):=\langle a,\beta\rangle$ and where
$\langle\cdot,\cdot\rangle$ denotes the natural pairing between $\g$ and
$\g^*$. The symplectic structure is then given by
$\t\omega=\omega+\hat\omega$. The left action $\hat\Phi$ of $T^*G$ 
on $\g^*\oplus\g$ is the canonical lift to the cotangent bundle 
of the affine action of $T^*G$ on the base $\g^*$:
$((g,\alpha),\beta)\rightarrow \Ad^*_{g^{-1}}(\beta)+\alpha$, so that 
for $\t\Phi$ we thus obtain
\begin{equation}
\label{A2}
\t\Phi_{(g,\alpha)}\left(m,\beta,b\right):=
\left(\Phi_g(m)\,,\,\Ad^*_{g^{-1}}(\beta)+\alpha\,,\,\Ad_g(b)\right).
\end{equation}
Since $\hat\Phi$ is a cotangent lift, it automatically preserves 
$\hat\omega$ and hence $\t\Phi$ preserves $\t\omega$. Also, by the same 
token, the momentum map $\hat J$ for $\hat\Phi$ is simply given by 
\begin{equation}
\label{A3}
\langle(a,\alpha),\hat J(\beta ,b)\rangle
=\sigma\left(\frac{d}{dt}\Bigg\vert_{t=0}
{\hat\Phi}_{(g(t),\alpha(t))}(\beta,b)\right)
=\langle b,-\ad^*_a(\beta)+\alpha\rangle ,
\end{equation}
where $(g(0),\alpha(0))=(e,0)$ and 
$(a,\alpha):=\frac{d}{dt}\big\vert_{t=0}(g(t),\alpha(t))$. 
Hence the total momentum map for $\t \Phi$ is defined by
\begin{equation}
\label{A4}
\langle(a,\alpha),\t J(m,\beta,b)\rangle
=\langle a,J(m)\rangle-\langle b,\ad^*_a(\beta)\rangle
+\langle a,\alpha\rangle .
\end{equation}
The extended constraint set ${\t J}^{-1}(0)\subset M\times \g^*\oplus\g$ 
is now seen to be given by $J^{-1}(0)\times \g^*\oplus\{0\}$, which 
contains the extra dimensions parameterized by $\g^*$.  However, (\ref{A2}) 
shows that the enlarged gauge group just cancels these additional (ghost-) 
degrees of freedom and that 
$T^*G\backslash {\t J}^{-1}(0)\cong G\backslash J^{-1}(0)$, i.e., 
${\t M}_{\rm red}\cong M_{\rm red}$.

\subsection{Unimodularisation in Geometric Quantization}

In geometric quantization, the Hilbert space is built by 
completion from certain smooth sections in a line bundle over 
the symplectic manifold $(M,\omega)$ of dimension, say, $n$. 
A central structural element is a foliation $P$ of $M$ by 
$n$-dimensional Lagrangian submanifolds such that $M/P$ 
is smooth. For simplicity we shall here restrict attention to the 
subclass of so-called `half-density quantizations', where the line 
bundles are of the form $L=\hat L\otimes\vert\Lambda^nP\vert^{\frac{1}{2}}$, 
where $\hat L$ is a line bundle whose curvature is proportional to $\omega$, 
and where $\vert\Lambda^nP\vert^{\frac{1}{2}}$ is the line bundle of 
half-densities associated to $P$. The Hilbert space is obtained by 
completion from smooth sections, which are covariantly constant along $P$ 
(they are `polarized') and compactly supported on $M/P$. 
For this we clearly also need to choose a Hermitian structure 
that is compatible with the connection, which is always possible. 
The quantizable observables are those whose Hamiltonian flows preserve 
$P$ and there is a quantization map that maps smooth functions on $M$ 
to differential operators on $L$.

If $(M,\omega,G,\Phi,J)$ is a gauge system as above, then, provided 
that $P$ is compatible with the action $\Phi$ of $G$ in the sense 
that $P$ is invariant under $\Phi$ and satisfies 
$TP\cap T^{\perp}(J^{-1})=\{0\}$, there is an induced polarization 
$P_{\rm red}$ in $(M_{\rm red},\omega_{\rm red})$. Furthermore, if $\Phi$ 
lifts to a connection preserving  action on $L$, we can perform a 
``$G$-invariant'' quantization based on $(M,\omega,P)$. The natural question 
then is whether this quantization is unitarily equivalent to the
already reduced quantization based on 
$(M_{\rm red},\omega_{\rm red},P_{\rm red})$. Here the main result 
is (see \cite{DE1}), that this is indeed the case if $G$ is 
unimodular and the condition of $G$-invariance on sections in L is 
just the Dirac condition (where $J_a:=\langle a,J\rangle$ and
$\J_a$ is its quantization)
\begin{equation}
\label{Dirac}
(\J_a)\psi=0\quad\forall a\in\g\,.
\end{equation}
Hence in this sense the Dirac condition may be thought of as being 
derived from the requirement of equivalence with the  
quantization based on the reduced phase space. However, in the 
proof unimodularity is needed to (naturally) relate half 
densities on $(M,P)$ to those on $(M_{\rm red},P_{\rm red})$ so that 
this result is not directly applicable to the non-unimodular case.
But the trick is of course to apply it to the unimodularized setting.
Then the Dirac condition for the $\t G$ invariant quantization of the 
extended system can be shown to the equivalent to the following modified 
Dirac condition for a $G$-invariant quantization of the original system:
\begin{equation}
\label{Diracmodinf}
(\J_a)\psi=\frac{i}{2}\hbox{tr}(\ad_a)\psi\quad\forall a\in\g\,.
\end{equation}
which is just the infinitesimal version of (\ref{Diracmod}) and which
reduces to (\ref{Dirac}) for unimodular $G$, since then
the $\ad_a$ are traceless.  Hence we conclude that in the non-unimodular 
case the physical Hilbert space is to be built from solutions of 
(\ref{Diracmodinf}) (as a differential equation, not as an eigenvalue equation
in Hilbert space) rather than (\ref{Dirac}).
The failure of the Dirac condition has also been explicitly studied 
in simple examples with non-unimodular gauge group, like the 
pseudo-rigid body~\cite{DE2} and other systems~\cite{Tu}\cite{DE1}.

\subsection{Unimodularisation and Group Averaging}

Let $G$ and $\t G=T^*G\cong G\ltimes\g^*$ be as above and
recall the multiplication law: 
$(g_1,\alpha_1)(g_2,\alpha_2)=(g_1g_2,\alpha_1+\Ad^*_{g_1^{-1}}(\alpha_2))$.
We shall identify $G$ and $\g^*$ with subgroups of $\t G$ as follows:
$G\ni g\leftrightarrow (g,0)\in\t G$ and
$\g^*\ni\alpha\leftrightarrow (e,\alpha)\in\t G$, where $e$ denotes
the identity in $G$. Since $\t g=(g,\alpha)=(e,\alpha)(g,0)$, this allows 
us to simply write $\t g=\alpha g$ 
and hence to uniquely decompose 
each element of $\t G$ into the product of elements from $\g^*$ and $g$ 
respectively. The multiplication of ${\t g}_1$ with ${\t g}_2$ can then 
simply be written as follows: 
\begin{equation}
\label{A5}
{\t g}_1{\t g}_2 = \alpha_1g_1\, \alpha_2g_2
                 = \alpha_1(g_1\alpha_2 g_1^{-1})\, g_1g_2\,, 
\end{equation}
where $g_1\alpha_2g_1^{-1}=\Ad^*_{g^{-1}_1}(\alpha_2)$, so that in
$\t G$, the 
automorphism $\Ad^*_{g^{-1}}$ of $\g^*$ is written as simple 
conjugation.

Since $\t G$ and $\g^*$ are unimodular, their invariant measures are 
necessarily bi-invariant. We now construct the bi-invariant measure 
$d\t g$ on $\t G$ as a product measure of the bi-invariant measure 
$d\alpha$ on $\g^*$ and the right-invariant measure $d_Rg$ on $G$. 
More precisely, if $\t g=\alpha g$, we claim: 
\begin{equation}
\label{A6}
d\t g=d\alpha\, d_Rg\,,
\end{equation}
where this equality should be understood as saying that there exists 
equality if the undetermined constant scale-factors in each measure 
are chosen appropriately. For the proof it suffices to show
left- and right-invariance under $\t G$-multiplication of the right hand side 
of (\ref{A6}). Left-Invariance follows from
\begin{eqnarray}
\label{A7}
d(\t g_1\t g) &=& d(\alpha_1+\Ad^*_{g_1^{-1}}(\alpha))\, d(g_1g) 
                 \nonumber\\
              &=& \det\{\Ad^*_{g_1^{-1}}\}\,d\alpha\ \Delta(g_1)\,dg
                 \nonumber\\ 
              &=& d\alpha\, dg=d\t g\,,                           
\end{eqnarray}
where we used (\ref{measures}) and $\Delta(g)=\det\{\Ad_g\}$.  
Right-Invariance is even simpler:
\begin{eqnarray}
\label{A8}
d(\t g\t g_1) &=& d(\alpha+\Ad^*_{g^{-1}}(\alpha_1))\, d_R(gg_1) 
                 \nonumber\\
              &=& d\alpha\, dg = d\t g\,.
\end{eqnarray}

Suppose now that the unimodular group $\t G$ acts on a Hilbert 
space ${\t{{\cal H}}}_{\rm aux}$ via an unitary representation $\t U$.
Suppose further that we have an associated rigging map $\t\eta$, given by 
averaging over $\t G$. We can then write (in the following all 
integrals are to be understood in the weak sense): 
\begin{eqnarray} \label{A9a}
\t\eta:=\int_{\t G} d\,\t g\t U(\t g)
&=& \int_{\g^*}d\alpha\,\t U(\alpha)\int_G d_Rg\,\t U(g) \\
\label{A9b}
&=& \int_G d_Lg\,\t U(g)\int_{\g^*}d\alpha\,\t U(\alpha) \\
\label{A9c}
&=:&\eta\circ\hat{\eta}\,,                                        
\end{eqnarray}
where we denoted the averaging maps over $G$ and $\g^*$ in (\ref{A9b})
by $\eta$ and $\hat\eta$. The equality of the first with the second 
line follows from $d_Lg=\Delta^{-1}(g)d_Rg$ and
\begin{eqnarray}
\label{A10}
\int_{\g^*}d\alpha\t U(\alpha)\t U(g)
&=& \t U(g)\int_{\g^*}d\alpha\,\t U(g^{-1}\alpha g)    \nonumber\\
&=& \t U(g)\int_{\g^*}d\alpha\,\t U(\Ad^*_g(\alpha))   \nonumber\\
&=& \Delta^{-1}(g)\t U(g)\int_{\g^*}\,d\alpha\t U(\alpha)\,, 
\end{eqnarray}
which we may write in the form
\begin{equation}
\label{A11}
\t U(g)\circ\hat{\eta}=\Delta(g)\,\hat{\eta}\circ\t U(g)\,.
\end{equation}

The point of the reordering in (\ref{A9b}) is that we wish to 
{\it first} reduce the normal subgroup $\g^*$. Since in our 
conventions the rigging-maps act to the right, we need to commute 
the $\g^*$-integration  to the right. At this intermediate step, 
i.e. after applying the rigging map connected with $\g^*$, 
we will then obtain an effective prescription for the rigging map 
$\eta$ connected with the remaining part $G$ of $\t G$. The task 
is to show that this prescription coincides with (\ref{GA}).

To verify this, we simply note that (\ref{A11}) leads to the statement 
that with respect to the `intermediate-inner-product', given by  
$\langle\phi_1,\phi_2\rangle_{\rm int}:=\hat{\eta}(\phi_1)[\phi_2]$, the 
maps $\t U(g)$ differ from being unitary by a factor of 
$\Delta^{\frac{1}{2}}(g)$. In fact, applying (\ref{A11}) to the second 
slot, we have
\begin{eqnarray}
\label{A12}
\langle\t U(g)\hat{\eta}(\phi_1),\t U(g)\hat{\eta}(\phi_2)\rangle_{\rm int}
&=& \Delta(g)\langle\t U(g)\hat{\eta}(\phi_1),\hat{\eta}(\t U(g)\phi_2)
   \rangle_{\rm int}                                   \nonumber\\
&=&\Delta(g)(\t U(g)\hat{\eta}(\phi_1))[\t U(g)\phi_2] \nonumber\\
&=&\Delta(g)\hat{\eta}(\phi_1)[\phi_2]                    \nonumber\\
&=&\Delta(g)\langle\hat{\eta}(\phi_1),\hat{\eta}(\phi_2)\rangle_{\rm int}\,,
\end{eqnarray}
where the third equality simply uses the fact that $\t U$ acts on the 
image of $\hat{\eta}$ via the dual action. Hence we see that the 
restriction of $\t U$ to $G\subset\t G$ can be written as a product of 
representations of $G$ 
\begin{equation}
\label{A13}
\t U\Big\vert_G=:\Delta^{\frac{1}{2}}\otimes U\,,
\end{equation}
where $U$ is unitary. Hence we can write 
the remaining integration over $G$, i.e. the $\eta$-map, as
\begin{equation}
\label{A14}
\eta=\int_Gd_Lg \ \t U(g)=\int_G{\d}g \ U(g)\,,
\end{equation}
which indeed coincides with (\ref{GA}).

\end{document}